\documentclass[aps,prl,onecolumn,superscriptaddress,groupedaddress]{revtex4}
\usepackage{subfig}
\usepackage{graphicx}  
\usepackage{dcolumn}   
\usepackage{bm}        
\usepackage{amssymb}   
\usepackage{slashed}
\usepackage{graphicx}				
\usepackage{amsmath}
\usepackage{mathtools}
\usepackage{tikz,pgf}
\usepackage{comment}
\usepackage{asymptote}
\usetikzlibrary{arrows,backgrounds}
\usetikzlibrary{fit,scopes,calc,matrix,positioning,decorations.pathmorphing}
\usepackage[all]{xy}
\usepackage{yfonts}
\newcommand{\bra}[1]{\ensuremath{\left\langle#1\right|}}
\newcommand{\ket}[1]{\ensuremath{\left|#1\right\rangle}}
\newcommand{\Bracket}[1]{\ensuremath{\left\langle#1\right\rangle}}

\begin{document}
\title{Uhlmann-Higgs stabilisation of modified causal structures}
\author{Andrei T. Patrascu}
\address{FAST Foundation, Destin FL, 32541, USA\\
email: andrei.patrascu.11@alumni.ucl.ac.uk}
\begin{abstract}
We present a dynamical gauge theory for purified quantum fields, based on the Uhlmann connection in the extended system–ancilla Hilbert space, and demonstrate that nontrivial Uhlmann holonomies induce global modifications of spacetime causal structure.  By coupling purified electromagnetic modes to an emergent U(n) gauge field, we derive an effective metric in the geometric‐optics limit whose curvature contributions tilt and stretch lightcones.  Stabilisation of these deformations is achieved via a Higgs‐like scalar in purification space, whose spontaneous symmetry breaking gives mass to the broken Uhlmann gauge directions and protects the resulting curvature.  A 1+1D toy model illustrates both static and time‐dependent holonomy zones and dynamical backreaction from multiple particles, providing a self‐consistent, emergent causal geometry.  Our framework reveals a novel intersection of mixed‐state geometric phases, open‐system dynamics, and emergent gravity‐like phenomena, and suggests photonic experiments for gauge‐induced lightcone engineering.
\end{abstract}
\maketitle
\section{Introduction}
The causal structure of spacetime underpins our understanding of relativistic physics, fixing the lightcones that delineate signal propagation and forbidding closed timelike curves in globally hyperbolic backgrounds. Yet, in quantum systems with intrinsic gauge degrees of freedom, emergent geometric phases can deform effective lightcones in ways that challenge classical intuition. In this work, we formulate a dynamical gauge theory for purified quantum fields based on the Uhlmann connection - a parallel transport in the enlarged Hilbert space of system plus ancilla - and show how nontrivial Uhlmann holonomies induce robust, global modifications of causal structure.

We begin by constructing an effective action for purified electromagnetic modes minimally coupled to an emergent U(n) gauge field, deriving a modified Hamilton–Jacobi equation in the geometric optics limit. The resulting effective metric acquires curvature contributions from the Uhlmann field strength, tilting and stretching the lightcones of semiclassical wavepackets. To stabilize these gauge‐induced deformations, we introduce a Higgs‐like scalar in purification space whose spontaneous symmetry breaking gives mass to broken components of the Uhlmann connection, thereby localizing and protecting its curvature. A 1+1D toy model illustrates both static and time‐varying holonomy zones, as well as backreaction from multiple particles, demonstrating a self‐consistent, emergent causal geometry.

Our results highlight a novel avenue by which quantum information structures - encoded in mixed‐state purifications - can manifest as physical modifications of spacetime causal boundaries. This Uhlmann gauge framework [1] bridges quantum open‐system dynamics [2], geometric phases of mixed states [3], and emergent gravity‐like phenomena, and opens new possibilities for experimental tests of gauge‐induced lightcone engineering in photonic platforms.
\section{open and closed systems}
The origin of spacetime metric can be seen as being in taking a semiclassical limit [4] while analysing quantum correlations between different modes [5]. We can in fact generate a metric and associate a notion of distance to correlation such that the metric would be inverse proportional to the specific correlation between modes before taking the semiclassical limit [6]. However, the description differs when we consider a system made out of a quantum fluid, for example a spin liquid, case in which the emergent metric and effective causal structure will only affect instruments that are embedded in the system and that use the degrees of freedom of that system as distance and time measuring devices [7], and the case in which we consider an open system, for example the coherent superposition of light, which would then in the semiclassical limit generate our usual spacetime causal structure [8]. In particular, if we want to engineer semiclassical limits of delicate and unstable correlations that would result in causal structures that would resemble exotic spacetimes like closed causal loops or extended causal cones, when dealing with a quantum fluid, those phenomena could be stabilised but would only be detectable through metrics and associated measuring devices attached to the interior degrees of freedom of the fluid. If however the same procedure is repeated with coherent photon mode superpositions, the emergent causal structure would reflect the causal structure of the region itself. 
When we say that the classical limit of a quantum fluid (e.g. a superfluid or a quantum Hall fluid) might permit closed causal loops, we typically mean that such causal anomalies (loop like behaviour) are localised to internal effective geometries, such as emergent spacetime geometries arising from collective excitations or topological structures (for example anyons in a 2D fluid). These effective causal structures are confined to the medium, emergent from microscopic degrees of freedom, like quasiparticles or spin-charge separation. These are not necessarily capable of influencing the ambient (external) spacetime geometry. The closed causal loops here are therefore artefacts of the material's excitation structure, not actual modification of overall causality [9]. 
We contrast that with the setup in which light is purified by means of an Uhlmann gauge theory and entangled with ancilla modes. Here the configuration allows potential large-scale modification of the causal structure. This occurs because photonic modes are not spatially confined. Unlike interior modes in a condensed matter system, purified classical electromagnetic field modes propagate through space and can carry global information that has been controlled by Uhlmann holonomies. These Uhlmann holonomies are the gauge data from parallel transport in the purified Hilbert space, an extension of the geometric phase space of Uhlmann that allows for dynamical degrees of freedom. The Uhlmann connections induce nontrivial geometric distortions akin to parallel transport over fibre bundles, but in the space of the purification of the classical electromagnetic wave. When light is purified and undergoes evolution through a non-trivial Uhlmann loop, its effective propagation phase and directionality can shift. This can lead to apparent faster than light effective motion (from a classical perspective) and a shifted causal boundary, as Uhlmann holonomy encodes a non-local quantum correction to how signals are traced back to sources. Within the Uhlmann dynamical gauge theory we can consider spontaneous symmetry breaking. This would introduce a vacuum alignment for the purified quantum state. It would set a preferred frame or direction in Hilbert space gauge geometry, which can then project back into modified spacetime geodesics. This projection alters the causal propagation cones of particles and fields, not just internally, but also in the bulk spacetime itself, particularly in regions where the purified field dominates. Therefore, while in quantum fluids causal anomalies remain internal, because the excitations are material bound and don't carry gauge induced geometric phase beyond the system, in the dynamical Uhlmann gauge structure, purified light fields are free space propagating and their holonomic structure feeds back into the causal metric, perceived by semiclassical observers. 
This is the main reason why such a setup would imply causal structure modification and not just an internal effect. 
Let us first consider a mixed photonic state described by a density matrix $\rho$. Through purification, we embed $\rho$ into a larger Hilbert space 
\begin{equation}
\ket{\Psi}\in \mathcal{H}_{sys}\otimes \mathcal{H}_{anc}
\end{equation}
The evolution of this purified state under a thermal or noisy channel $\epsilon$ can be treated as parallel transport with respect to the Uhlmann connection $A_{\mu}$. This defines a gauge theory over the space of purifications. The Uhlmann holonomy appears when we transport the purified state around a closed loop in parameter space 
\begin{equation}
\Gamma_{U}=\mathcal{P}exp(\oint A_{\mu}dx^{\mu})
\end{equation}
This holonomy modifies the phase of the purified state in a nontrivial geometric way, analogous to a Wilson loop in non-Abelian gauge theory. We next derive an effective action of the photonic field interacting with the Uhlmann background. As opposed to my previous papers, where I discussed the effects only in the Uhlmann bundle of purification and I modified explicitly the phase there in order to achieve the causal modification effects in the spacetime geometry, I will apply here a completely equivalent method but where the focus is on spacetime causal structure. I do this in order to perform certain calculations in a more suggestive manner. Let me therefore consider the Lagrangian for a purified electromagnetic field $A_{\mu}$ now minimally coupled to an emergent Uhlmann gauge potential $\mathcal{A}_{\mu}$ now constructed as a projection in spacetime. 
\begin{equation}
\mathcal{L}_{eff}=-\frac{1}{4}F^{\mu\nu}F_{\mu\nu}+i\bar{\psi}\gamma^{\mu}(\partial_{\mu}+i\mathcal{A}_{\mu})\psi+...
\end{equation}
Here $\psi$ is an effective spinor encoding the quantum degree of freedom for purification and $\mathcal{A}_{\mu}$ captures the Uhlmann connection projected into spacetime. It is worth noting that we included a spinor Ansatz for convenience, as a simple way to encode the extra "ancilla" degrees of freedom that live alongside each photon mode in the purification. Any field carrying a representation of U(n) would do. We chose to write the purified photonic mode as a Dirac like spinor $\psi(x)$ only because it is the most familiar way to couple matter to a gauge connection via a covariant derivative $D_{\mu}=\partial_{\mu}+i\mathcal{A}_{\mu}$. In reality the purified mode could be a multiplet of scalars, vectors, or higher spin fields, what matters is that it transforms under the Uhlmann gauge group. Spinors are helpful however because with a spinor $\psi$ the term $\bar{\psi}\gamma^{\mu}D_{\mu}\psi$ is the standard way to see how $\mathcal{A}_{\mu}$ modifies the kinetic propagation and then in the eikonal limit  the $\gamma^{\mu}$ picks out the effective metric deformation. In the geometric optics limit one writes $\psi\sim a\cdot e^{iS/\hbar}$. The ensuing Hamilton Jacobi equation $\bar{\psi}\gamma^{\mu}\psi\partial_{\mu}S=0$ generalises to $g_{eff}^{\mu\nu}\partial_{\mu}S\cdot \partial_{\nu}S=0$. We can connect back to pure light and trade the spinor for a two-component complex field $(\psi_{1}, \psi_{2})$ or even a single scalar doublet, carrying the same U(n) indices. The physics of Uhlmann holonomy induced curvature and the Higgs mechanism is representation-agnostic. We just used the spinor to keep the analogy with the standard gauge theory and to make the transition to a lightcone deformation as visible as possible. 
In any case, this coupling modifies the dispersion relations of photons through geometric phase accumulation, analogous to Berry curvature corrections in solids but now in spacetime. 
In the eikonal approximation of geometric optics, we write the field as 
\begin{equation}
A_{\mu}(x)=\epsilon_{\mu}(x)e^{iS(x)/\hbar}
\end{equation}
We then substitute into the wave equation and keep only leading terms in $\hbar$, and we obtain the modified Hamilton Jacobi equation 
\begin{equation}
g_{eff}^{\mu\nu}(x)\partial_{\mu}S\cdot \partial_{\nu}S=0
\end{equation}
where $g_{eff}^{\mu\nu}$ is the effective metric experienced by the photonic wavepacked. It gets corrected as
\begin{equation}
g_{eff}^{\mu\nu}=\eta^{\mu\nu}+\alpha\mathcal{F}^{\mu\nu}+\beta\nabla^{(\mu}\mathcal{A}^{\nu)}+...
\end{equation}
with 
\begin{equation}
\mathcal{F}^{\mu\nu}=\partial^{\mu}\mathcal{A}^{\nu}-\partial^{\nu}\mathcal{A}^{\mu}
\end{equation}
where $\alpha$ and $\beta$ are model-dependent coefficients. These corrections deform the causal structure: the lightcone becomes tilted or stretched, allowing superluminal effective propagation or non-local signal redirection within the consistency of the purified gauge geometry. This modified $g_{eff}^{\mu\nu}$ defines a new causal boundary. Signals that were once outside the forward lightcone in Minkowski space can now fall inside the new, wider cone, as governed by 
\begin{equation}
ds_{eff}^{2}=g_{\mu\nu}^{eff}dx^{\mu}dx^{\nu}<0
\end{equation}
That is, events previously causally disconnected in flat space can now be effectively connected due to the influence of the Uhlmann connection. 
The Uhlmann gauge field is not physical in spacetime but its effect projects onto spacetime dynamics via the purification. This is a gauge-induced shift in perceived geometry, not a violation of relativity, more related to the idea of emergent spacetime from entanglement. 
The light appears to move faster because the purification alters the geodesic structure, not because it literally exceeds c in the original Minkowski background. 
\section{Path integral formulation: Uhlmann Holonomy as a Boundary Term}
We may begin with the partition function for a quantum system coupled to an environment purified via ancilla degrees of freedom. The path integral over purified field configurations is 
\begin{equation}
Z=\int \mathcal{D}[\Psi,\bar{\Psi}]e^{i S_{eff}[\Psi,\bar{\Psi},\mathcal{A}_{\mu}]}
\end{equation}
where $\Psi$ is the purified state evolving in an extended Hilbert space and $\mathcal{A}_{\mu}$ is the Uhlmann gauge field arising from the parallel transport condition
\begin{equation}
\Bracket{\Psi(x)|\Psi(x+dx)}=max
\end{equation}
This implies minimal distance between neighbouring purifications in the Bures metric, giving rise to a connection $\mathcal{A}_{\mu}$ such that 
\begin{equation}
\delta S=\int_{\partial M}\bra{\Psi}\mathcal{A}_{\mu}\ket{\Psi}dx^{\mu}
\end{equation}
Hence the effective action picks up a boundary term 
\begin{equation}
S_{eff}=S_{0}+\oint_{\partial M}Tr(\rho\mathcal{A})dx^{\mu}=S_{0}+\phi_{u}
\end{equation}
Here $\phi_{U}$ is the Uhlmann phase. This is conceptually similar to the Berry phase term in topological systems, but generalised to mixed states via purification. 
This boundary term alters the stationary phase paths of the semiclassical approximation. The classical trajectories (e.g. of a photon wavepachet) are now geodesics in a geometry twisted by the holonomy encoded in $\phi_{U}$. 
\section{Stabilisation via Spontaneous Symmetry Breaking in the Uhlmann sector}
The purified state lives in a projective Hilbert space with a gauge symmetry under $U(n)$, if the density matrix is $n$-dimensional. The Uhlmann parallel transport condition defines a horizontal lift with a gauge field $\mathcal{A}_{\mu}\in u(n)$. The symmetry breaking pattern can look like 
\begin{equation}
U(n)\rightarrow H\subset U(n)
\end{equation}
where $H$ is a stabiliser subgroup of the purification, i.e. symmetries preserved by the ground purified state. We assume the dynamics in the Uhlmann sector induce a potential $V(\Phi)$ for a purification dependent Higgs like field $\Phi$, living in some representation of $U(n)$. A symmetry-breaking vacuum expectation value 
\begin{equation}
\Bracket{\Phi}\neq 0
\end{equation}
produces mass terms for some components of $\mathcal{A}_{\mu}$ via the usual Higgs mechanism
\begin{equation}
\mathcal{L}_{mass}\sim |\Bracket{\Phi}|^{2}\mathcal{A}_{\mu}^{2}
\end{equation}
This localises the gauge connection in the sense that the nontrivial holonomies become rigid, topologically protected and stable under fluctuations. With $\mathcal{A}_{\mu}$ stabilised via spontaneous symmetry breaking, the effective metric correction 
\begin{equation}
g_{eff}^{\mu\nu}\sim\eta^{\mu\nu}+f(\Bracket{\mathcal{F}})
\end{equation}
is now robust. It no longer fluctuates but instead becomes settled into a semiclassically consistent geometry. 
The effective lightcones derived from the modified Hamilton-Jacobi equations are thus globally coherent and not gauge-drifted, insensitive to noise in the purification dynamics, and physically testable, since the geometric effects manifest in real space photonic behaviour (phase shifts, time delays, etc.) 
We construct now a toy model in which we display the purified field (light mode and ancilla), the Uhlmann connection and its holonomy, and how the spontaneous symmetry breaking localises the gauge field. We also perform calculations resulting in the effective lightcone tilt by Uhlmann curvature. We consider a scalar model where light is encoded by a purified scalar field $\psi(x^{\mu})$ in $1+1$ dimensional Minkowski space $x^{\mu}=(t,x)$ minimally coupled to a Uhlmann gauge field $\mathcal{A}_{\mu}$. The field content is 
$\psi(x)$ the complex purified field (system plus ancilla), $\mathcal{A}_{\mu}(x)$ being the U(1) Uhlmann gauge field, $\phi(x)$ the real scalar Higgs like field. The action will be 
\begin{equation}
S=\int d^{2}[-|D_{\mu}\psi|^{2}-V(\psi)-\frac{1}{4}\mathcal{F}_{\mu\nu}\mathcal{F}^{\mu\nu}+\frac{1}{2}(\partial_{\mu}\phi)^{2}-U(\phi)+\lambda\phi^{2}\mathcal{A}_{\mu}\mathcal{A}^{\mu}]
\end{equation}
where $D_{\mu}\psi=(\partial+i\mathcal{A})\psi$, $\mathcal{F}_{\mu\nu}=\partial_{\mu}\mathcal{A}_{\nu}-\partial_{\nu}\mathcal{A}_{\mu}$, $U(\phi)=\frac{\mu^{2}}{2}\phi^{2}+\frac{\lambda'}{4}\phi^{4}$ is the spontaneous symmetry breaking for $\mu^{2}<0$. 
The potential $U(\phi)$ has a minimum at 
\begin{equation}
\Bracket{\phi}=v=\sqrt{-\mu/\lambda'}
\end{equation}
If we put now this back into the action we obtain the Uhlmann gauge field with a mass term 
\begin{equation}
\mathcal{L}_{mass}=\lambda v^{2}\mathcal{A}_{\mu}\mathcal{A}^{\mu}
\end{equation}
The fluctuations in $\mathcal{A}$ are now suppressed and the gauge field is localised, which makes the holonomy stable. 
Now, the causal structure in the geometric optic limit is obtained by taking the WKB limit 
\begin{equation}
\psi(x)=a(x)e^{i S(x)/\hbar}
\end{equation}
We insert this into the action and obtain to leading order in the kinetic term 
\begin{equation}
(D_{\mu}\psi)^{*}D^{\mu}\psi\sim |a|^{2}(\partial_{\mu}S+\mathcal{A}_{\mu})(\partial^{\mu}S+\mathcal{A}^{\mu})
\end{equation}
The stationary phase condition (in semiclassical propagation) is given by 
\begin{equation}
g_{eff}^{\mu\nu}(x)(\partial_{\mu}S+\mathcal{A}_{\mu})(\partial_{\nu}S+\mathcal{A}_{\nu})=0
\end{equation}
The effective inverse metric is 
\begin{equation}
g_{eff}^{\mu\nu}(x)=\eta^{\mu\nu}+\delta g^{\mu\nu}(x),\;\; \delta g^{\mu\nu}\sim f(\mathcal{F}^{\mu\nu})
\end{equation}
In 1+1 dimensions there is only one independent component 
\begin{equation}
\mathcal{F}^{01}=\partial_{t}\mathcal{A}_{x}-\partial_{x}\mathcal{A}_{t}
\end{equation}
so if we impose a static holonomy configuration, say
\begin{equation}
\mathcal{A}_{t}=0,\;\; \mathcal{A}_{x}=\frac{\theta}{L}
\end{equation}
then
\begin{equation}
\delta g^{\mu\nu}\sim \alpha\begin{pmatrix} 0 &1\\ 1 &0\end{pmatrix}\Rightarrow g_{eff}^{\mu\nu}=\begin{pmatrix} -1 & \alpha\\ \alpha & 1\end{pmatrix}
\end{equation}
This deforms the lightcone.
The Uhlmann gauge connection, while initially defined in the enlarged Hilbert space (system+ancilla), ultimately enters the physical equations via the effective metric experienced by semiclassical fields propagating in spacetime. The Uhlmann gauge connection $\mathcal{A}_{\mu}$ emerges as a non-Abelian connection on the space of purifications with curvature $\mathcal{F}_{\mu\nu}$ which is inherently antisymmetric. However when translating these gauge induced effects into observable geometric corrections like the effective metric, we take into consideration that the metric is symmetric. The gauge field initially acts like a geometric phase in purification phase, which does not directly prescribe spacetime geometry. Instead, the physical geometry arises from a projection of the gauge structure onto spacetime. In this projection, we decompose the effects of the Uhlmann gauge structure into symmetric and anti-symmetric contributions. Physically, antisymmetric contributions correspond to torsion effects which we ignore here and hence take only the symmetric contributions to the metric. However, given string-theoretical Kalb-Ramond terms as well as the possibility of torsion contributions in modified gravity theories and alternatives to dark matter, I found it pertinent to point out this approximation made and to remind the reader that such anti-symmetric contributions to the metric do arise in the Uhlmann informational approach. 
The modified null condition is 
\begin{equation}
g_{eff}^{\mu\nu}k_{\mu}k_{\nu}=0 \Rightarrow -k_{t}^{2}+2\alpha k_{t}k_{x}+k_{x}^{2}=0
\end{equation}
and solving this gives 
\begin{equation}
\frac{dt}{dx}=\frac{k_{t}}{k_{x}}=\alpha\pm\sqrt{1+\alpha}
\end{equation}
Therefore the group velocity shifts and the causal structure is tilted in the $(t,x)$ plane. The deformation is globally induced by the stable Uhlmann field. To mimic a localised Uhlmann loop, we could make 
\begin{equation}
\mathcal{A}_{x}(x)=\frac{\theta}{L}\chi_{[x_{1},x_{2}]}(x)
\end{equation}
where $\chi$ is the characteristic function. Then light only experiences a geometric twist between $x_{2}$ and $x_{2}$ like passing through a holonomy-active region. Outside, the lightcone is normal, inside it is distorted leading to nontrivial signal delay or advancement. 
Figure 1 shows how the lightcones are modified inside the Uhlmann holonomy zone (considered for this example between x=-2 and x=2). Outside this region the lightcones follow the normal 45 degrees slope, inside however the causal structure tilts and stretches allowing for effective superluminal propagation. The shift is symmetric due to the positive and negative branches of $dt/dx$ showing both forward and backward propagation are affected. 
\begin{figure}[h!]
 \centering
  \includegraphics[width=0.5\textwidth]{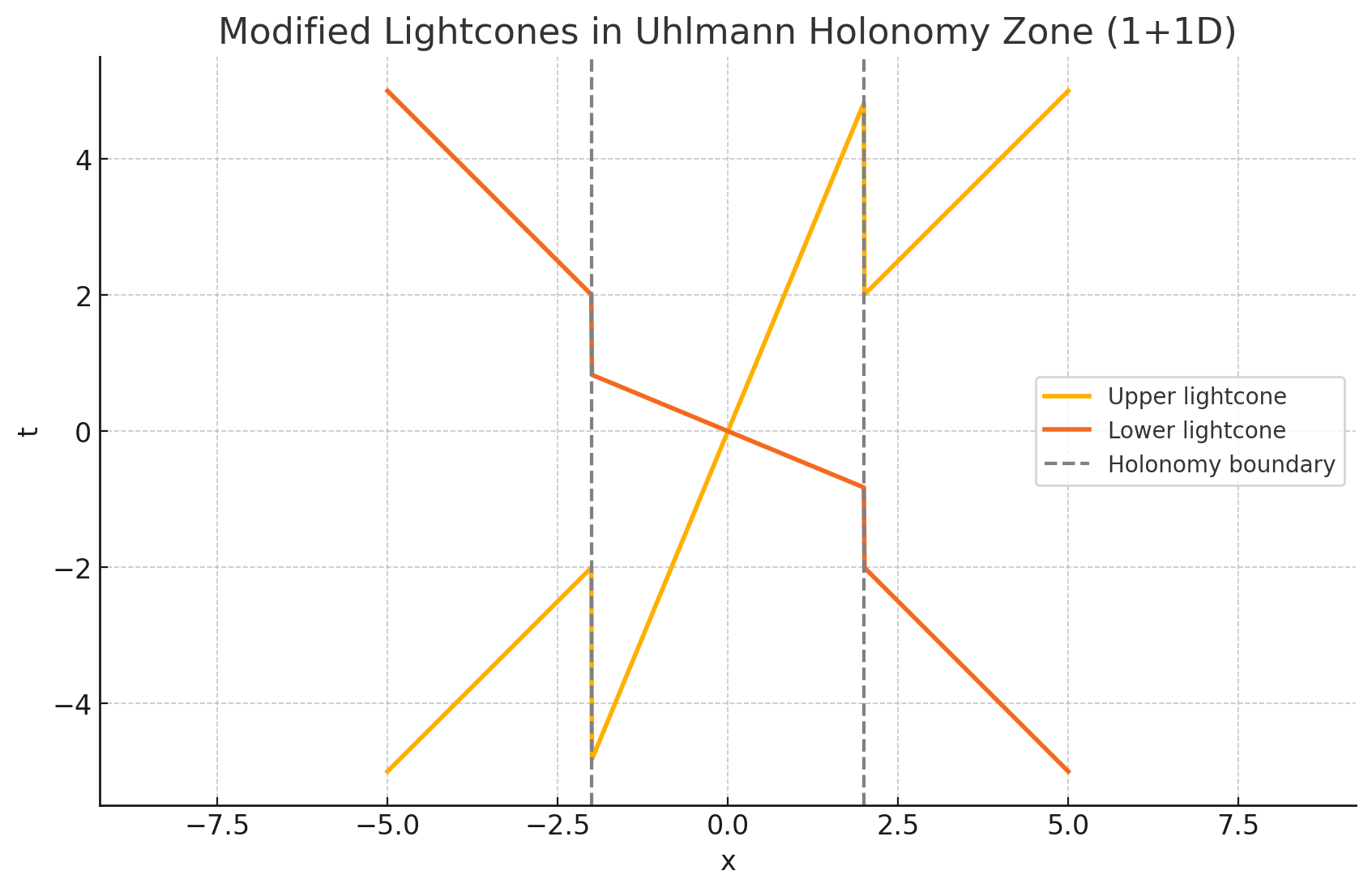}
 \caption{}
 \end{figure}
 In figure 2 we represent the time-varying lightcone tilting where the lightcone slope changes with both position and time. Inside the holonomy active region the Uhlmann induced curvature fluctuates, leading to a dynamical causal structure. Outside this zone, the geometry remains flat. 
 \begin{figure}[h!]
 \centering
  \includegraphics[width=0.5\textwidth]{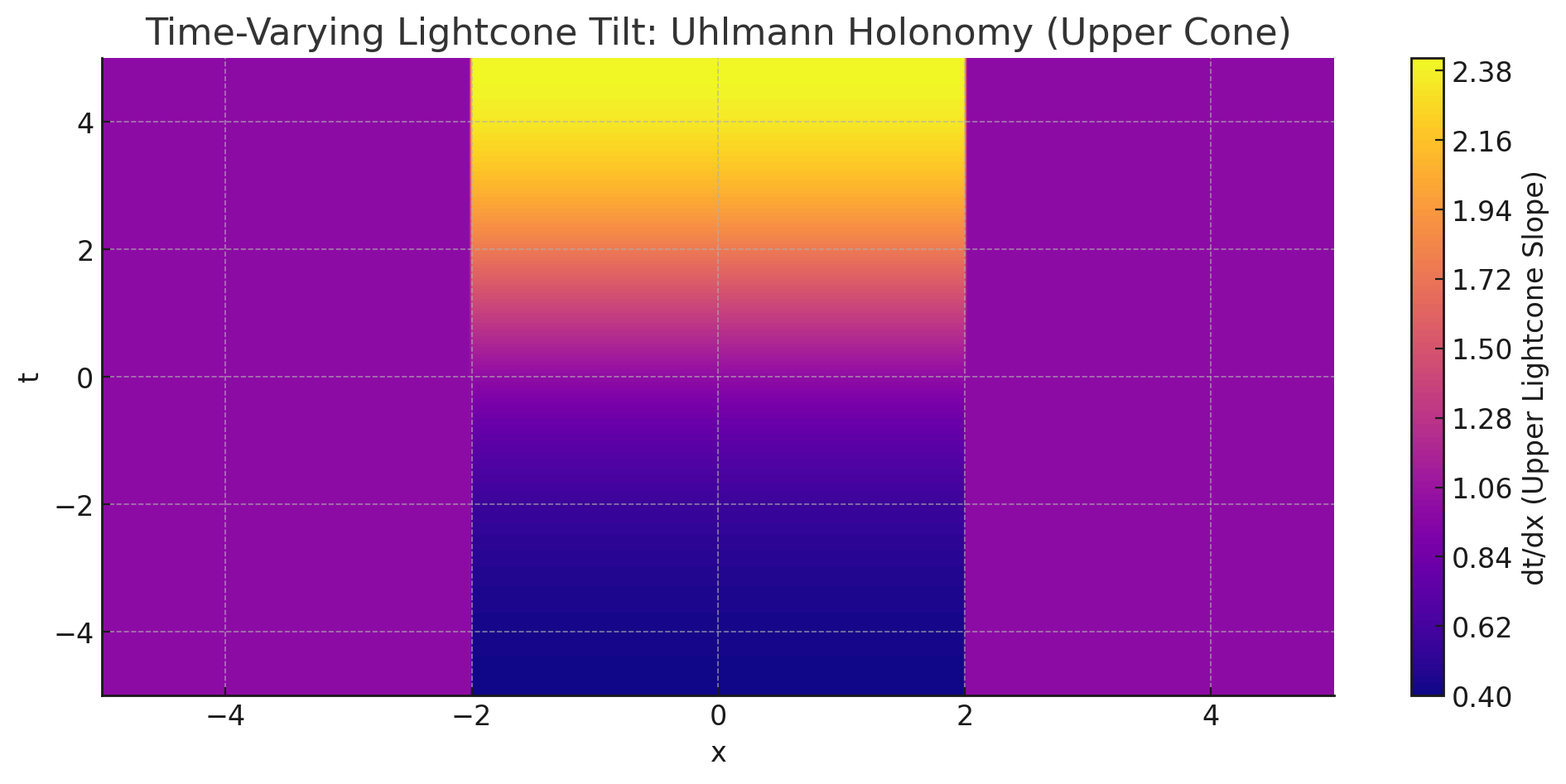}
 \caption{}
 \end{figure}

In figure 3, we depict a particle trajectory under the effective metric and obtain a simulated geodesic for a particle starting at $x=0$ and $t=0$. As the particle enters the holonomy zone, it experiences a tilted causal structure, speeding up its evolution. This shift is smooth and driven entirely by the time-dependent Uhlmann holonomy. 

\begin{figure}[h!]
 \centering
  \includegraphics[width=0.5\textwidth]{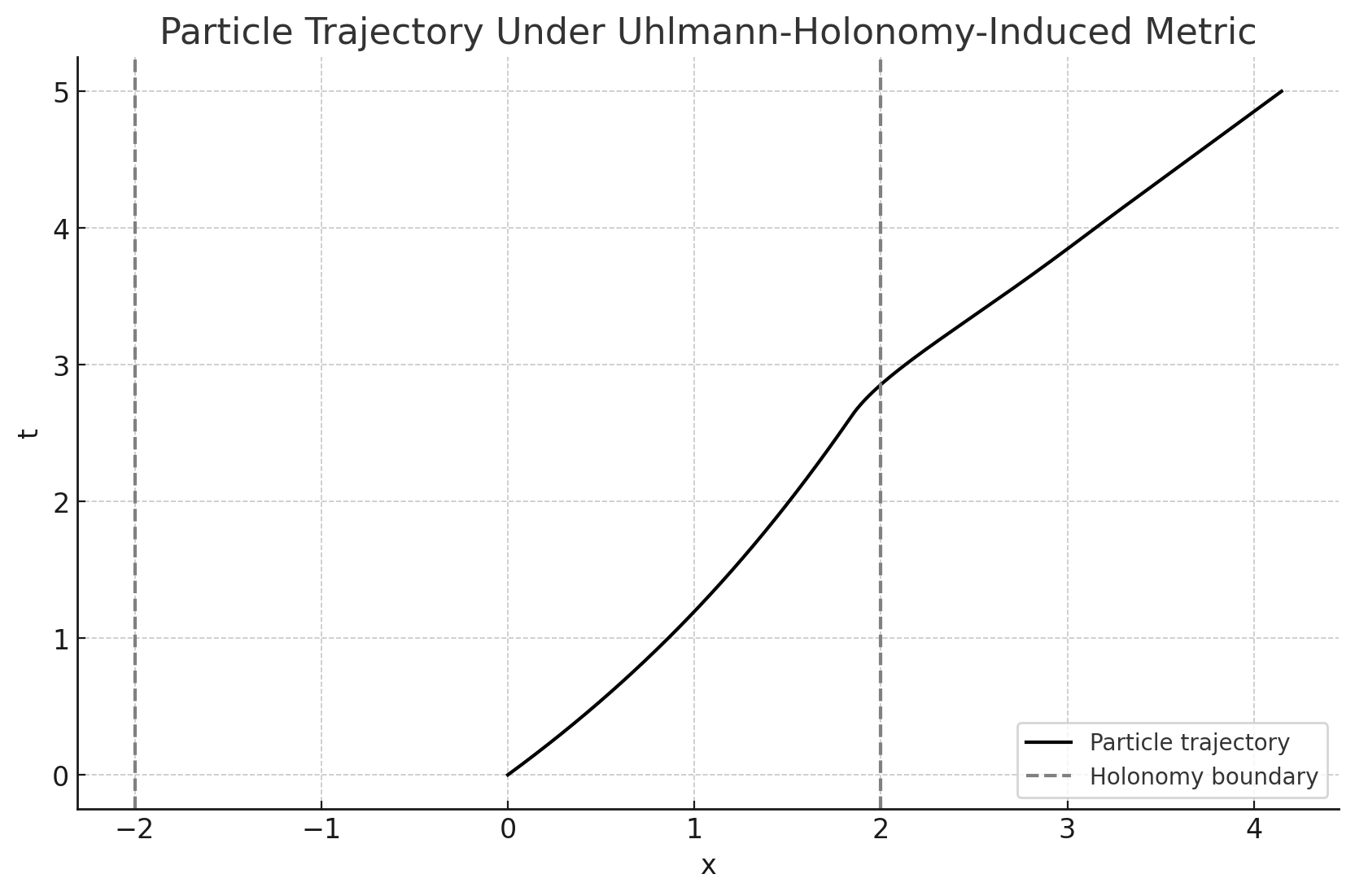}
 \caption{}
 \end{figure}

In figure 4 we introduce also backreaction, i.e. the particle affecting the holonomy. Each particle begins at a different position (-1.5, 0, 1.5) within the holonomy-active zone. As they move, their presence locally enhances the Uhlmann connection via a Gaussian shaped backreaction. This makes the effective lightcone even more tilted in regions where multiple particles overlap, resulting in varying accelerations across their paths. 
This mimics how matter can locally curve the effective causal geometry, via interactions with the underlying Uhlmann gauge structure, a toy model analogue to backreaction in general relativity. 
\begin{figure}[h!]
 \centering
  \includegraphics[width=0.5\textwidth]{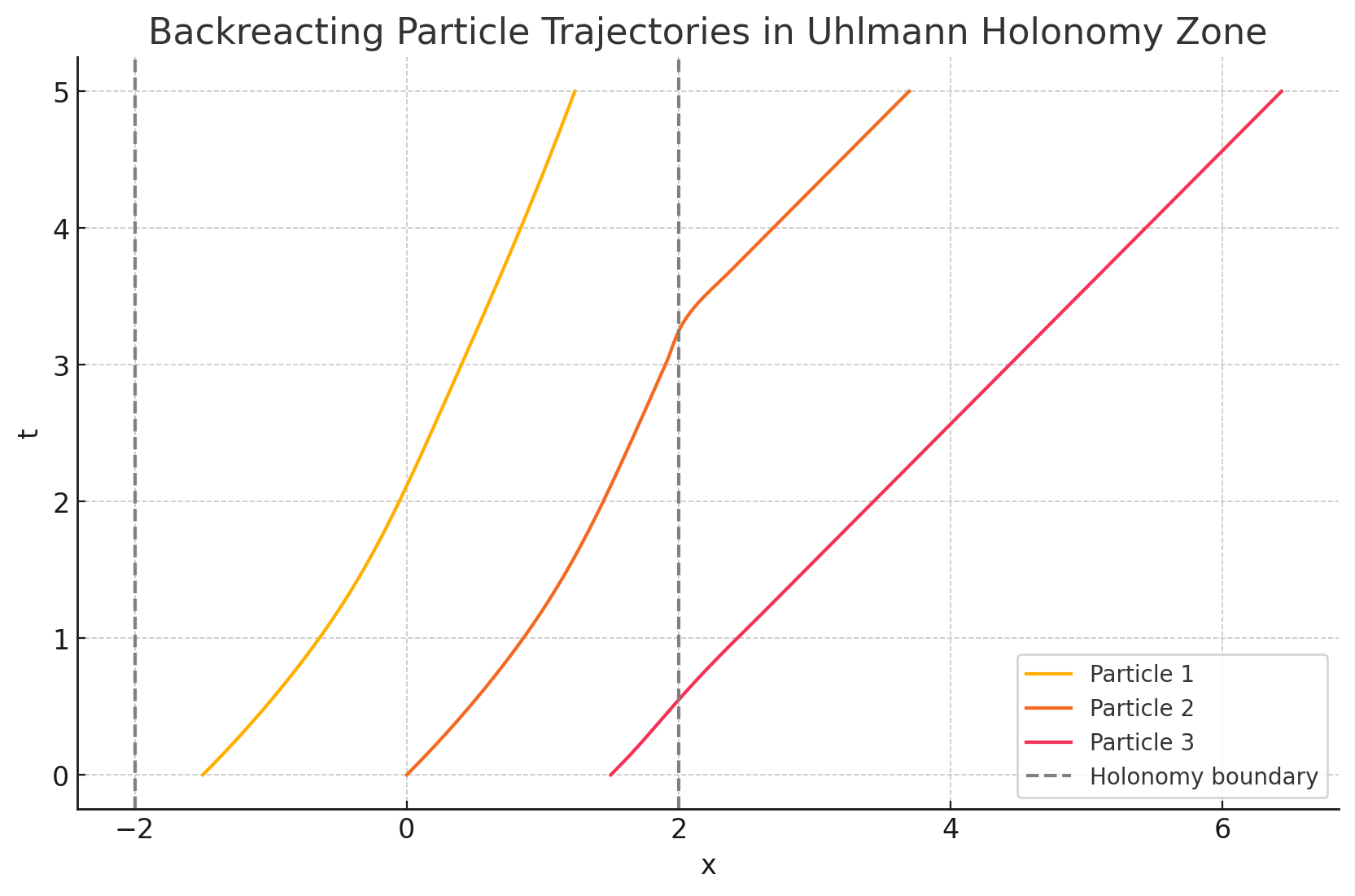}
 \caption{}
 \end{figure}
Finally, in figure 5 each particle's trajectory updates based on the instantaneous position of all other particles at each time step. They weakly accelerate or decelerate relative to each other, depending on how they concentrate within the holonomy zone. The deformation of the effective causal structure is now emergent, arising dynamically from the collective backreaction. This captures a simplified version of dynamical causal structure, where field excitations (particles) help shape the very geometry they move in, reminiscent of semiclassical gravity but now by means of the Uhlmann gauge field. 
\begin{figure}[h!]
 \centering
  \includegraphics[width=0.5\textwidth]{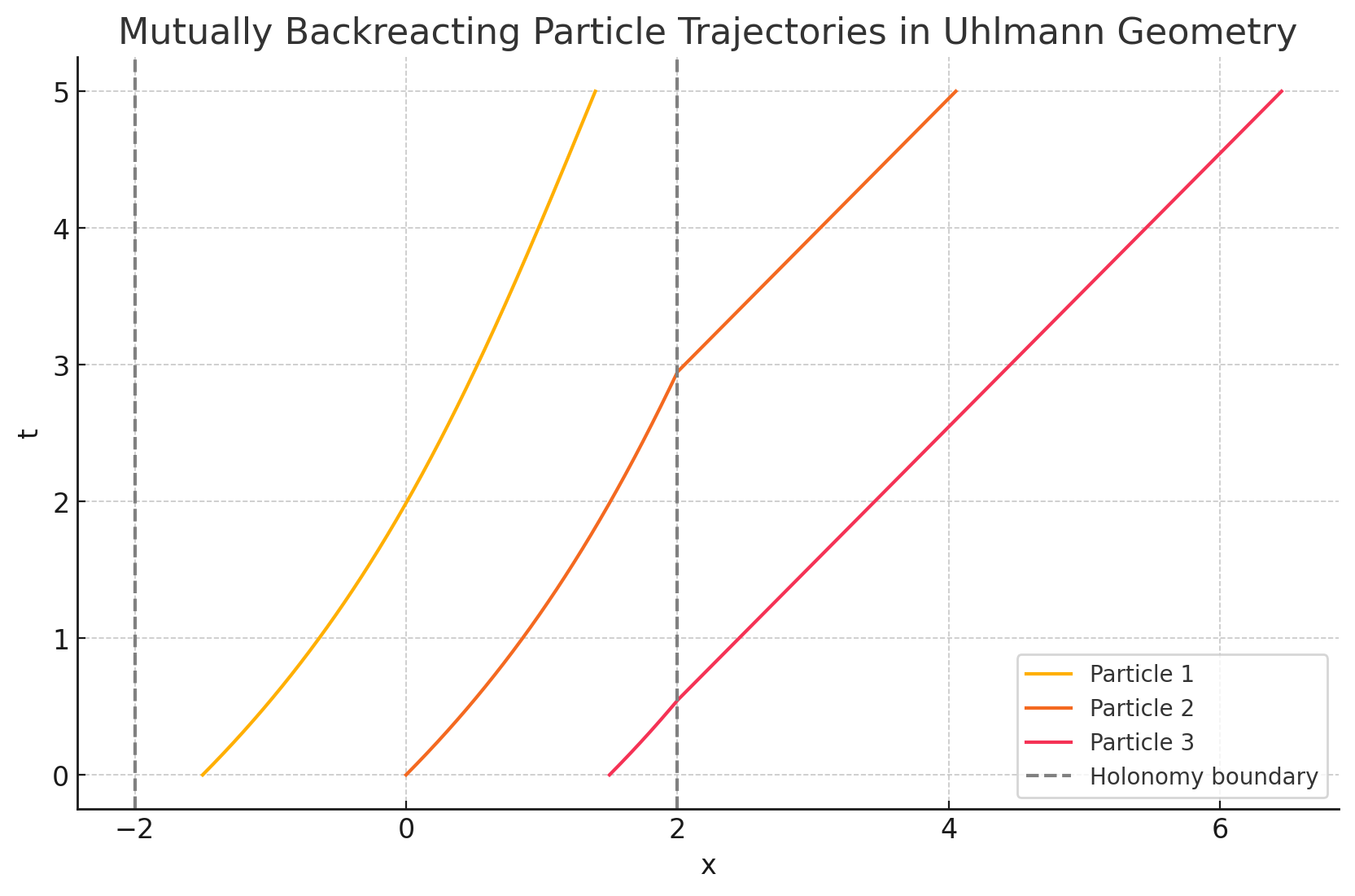}
 \caption{}
 \end{figure}

\end{document}